\documentclass[aps,prl,twocolumn,superscriptaddress,showpacs]{revtex4-1}

\usepackage[english]{babel}
\usepackage{graphicx}
\usepackage{amsmath,amssymb,amstext,amsfonts}
\usepackage{xfrac}

\graphicspath{{./}{./figs/}{./plots/}}
\DeclareGraphicsExtensions{.png,.jpg,.pdf,.eps}

\begin{document}

\title{High-Quality Electron Beams from Beam-Driven Plasma Accelerators by Wakefield-Induced Ionization Injection}

\author{A. Martinez de la Ossa}
\affiliation{Deutsches Elektronen-Synchrotron DESY, D-22607 Hamburg, Germany}
\author{J. Grebenyuk}
\affiliation{Deutsches Elektronen-Synchrotron DESY, D-22607 Hamburg, Germany}
\author{T. Mehrling}
\affiliation{Deutsches Elektronen-Synchrotron DESY, D-22607 Hamburg, Germany}
\author{L. Schaper}
\affiliation{Institut f\"{u}r Experimentalphysik, Universit\"{a}t Hamburg, D-22761 Hamburg, Germany}
\author{J. Osterhoff}
\affiliation{Deutsches Elektronen-Synchrotron DESY, D-22607 Hamburg, Germany}

\date{\today}

\begin{abstract}
We propose a new and simple strategy for controlled ionization-induced trapping of electrons 
in a beam-driven plasma accelerator. 
The presented method directly exploits electric wakefields to ionize electrons 
from a dopant gas and capture them into a well-defined volume of the accelerating 
and focusing wake phase, leading to high-quality witness-bunches. 
This injection principle is explained by example of three-dimensional particle-in-cell (PIC) 
calculations using the code OSIRIS. 
In these simulations a high-current-density electron-beam driver excites plasma waves 
in the blow-out regime inside a fully-ionized hydrogen plasma of density 
$5\times10^{17}~\mathrm{cm^{-3}}$. 
Within an embedded $100~\mathrm{\mu m}$ long plasma column contaminated with neutral helium gas, 
the wakefields trigger ionization, trapping of a defined fraction of the released electrons, 
and subsequent acceleration. 
The hereby generated electron beam features a $1.5~\mathrm{kA}$ peak current, 
$1.5~\mathrm{\mu m}$ transverse normalized emittance, 
an uncorrelated energy spread of $0.3\%$ on a GeV-energy scale, and few femtosecond bunch length.
\end{abstract}


\maketitle

Over the last decade, the field of plasma-wakefield acceleration of electrons 
with ultra-high field gradients surpassing $10~\mathrm{GV/m}$ has progressed steadily and rapidly. 
This is testified by the qualitative improvement of the accelerated beams 
during this period. 
In particular laser-driven wakefield accelerators~\cite{Tajima:1979bn} were improved significantly. 
Milestones, such as the realization of quasi-monoenergetic electron spectra~\cite{Mangles:2004uq,Geddes:2004fk,Faure:2004fk}, 
GeV-class beams~\cite{Leemans:2006dx}, enhanced stability~\cite{Osterhoff:2008zz,Hafz:2008kx}, 
controlled injection techniques for tunability~\cite{Faure:2006vn,Pak:2010zza,Gonsalves:2011fk}, 
and the application of the generated beams to drive compact XUV~\cite{Fuchs:2009uq} and X-ray sources~\cite{Kneip:2010kx}  
promoted plasma-based acceleration to a promising technique for future accelerators.

Meanwhile, beam-driven plasma wakefield acceleration (PWFA)~\cite{veksler:1956,Chen:1985pwfa} 
made great advancements, culminating in the demonstration of energy-doubling of 
part of the $42~\mathrm{GeV}$ SLAC electron beam~\cite{Blumenfeld:2007ph} in a distance of less than a meter.
However, despite this remarkable progress, the quality of electron bunches extracted from beam-driven schemes 
lags behind those obtained from laser-driven plasma accelerators. 
This may mainly be attributed to the so far insufficient control 
over the electron-injection process in PWFA, 
which has a fundamental impact on the initial beam phase-space population and, thus, on the final beam quality. 

Several controlled injection techniques for PWFA have been proposed but have not been 
experimentally verified yet, such as the external injection of a tailored witness 
beam~\cite{Hogan:FACET}, 
magnetically induced injection~\cite{Vieira:2011qn}, 
and laser-triggered ionization injection~\cite{Hidding:2012th,Li:2003}.
All these methods demand several elements in the experiment to act in concert 
to achieve injection into the appropriate wake region,   
e.g.~fs synchronization and $\mu \mathrm{m}$ alignment of a laser to the particle beam or 
the generation of an adequate witness beam and its matching into the plasma wakefield~\cite{Mehrling:PRSTAB}. 
These measures can be technically challenging to implement and a source of instabilities and, 
hence, may hamper the generation of high-quality electron bunches. 

An easier approach constitutes the injection of electrons in plasma 
by means of field-induced ionization of a dopant gas with appropriate ionization potential, 
e.g.~helium (He). 
It was recently discovered that this process can be initiated by the radial electric field 
of the driving beam~\cite{Oz:2007mn}. 
However, the lack of control over the event and its sensitivity on initial conditions 
of the driving beam micro-structure did not lead to the production 
of qualitatively interesting beams. 
In this work, we propose a new and straightforward strategy for 
controlling ionization injection of electrons into beam-driven plasma wakes, 
utilizing the wake electric fields only and, thus, providing improved beam quality.
This technique exploits the difference in absolute electric-field strength
in the blow-out regime~\cite{Rosenzweig:1991yx,Lotov:2004zz,Lu.PRL.96.165002} existing
between the accelerating and decelerating regions within the first wakefield bucket
to selectively ionize a small volume of a background dopant gas near the phase of 
maximum acceleration only. 
In this way the production of high-quality, ultra-short ($\sim\mathrm{fs}$), 
low-emittance ($\sim\mathrm{\mu m}$), multi-$\mathrm{GeV}$-energy electron beams 
from a relatively simple experimental setup is made possible.

PWFA in the blow-out regime uses a relativistic charged particle beam, 
which is short compared to the plasma wavelength and of higher density than the background plasma.
This driver beam expels plasma electrons from its high-density core, 
forming a co-propagating ion cavity.
The electric fields in this cavity or bubble may exceed the cold non-relativistic wave-breaking 
field $E_0 = (m c^2 / e)~k_p$, where $k_p = \sqrt{n_0 e^2 / \epsilon_0 m c^2}$ is 
the plasma wave number, $n_0$ the plasma particle density, 
$\epsilon_0$ is the vacuum permittivity, 
$c$ the speed of light, and $m$ and $e$ are the electron mass and charge, respectively.   
Current accelerator facilities provide short (rms lengths of $10-50~\mathrm{\mu m}$) 
and dense (currents of $1-25~\mathrm{kA}$) electron beams which are suitable to operate 
wakefields in the blow-out regime with accelerating fields of $\sim 100~\mathrm{GV/m}$ 
in plasmas with densities on the order of $10^{17}\mathrm{cm^{-3}}$. 
The amplitude of these accelerating electric fields is sufficient
for ionization of electrons from a high-ionization 
potential atomic species such as He, and their trapping into 
a well defined phase of the wake near the back of the ion cavity.
The ionization process caused by static (or slowly varying) electric fields of a magnitude 
sufficient to significantly deform the atomic potential barrier of an atom 
can be described by a tunneling probability~\cite{perlomov:66}, 
and has been determined for a number of atomic species~\cite{Ammosov:1986adk}. 
Writing the tunneling-ionization rate in an engineering formula yields~\cite{Bruhwiler:2003zz} 
\begin{equation}
\label{eq:ionization}
\begin{split}
W_{ADK}[\mathrm{fs}^{-1}] &\approx 1.52 \frac{4^{n^*} \xi_i[\mathrm{eV}]}{n^* \Gamma(2n^*)}\left(20.5\frac{\xi_i^{3/2}[\mathrm{eV}]}{E[\mathrm{GV/m}]}\right)^{2n^*-1} \nonumber \\
   & \times\exp{\left(-6.83\frac{\xi_i^{3/2}[\mathrm{eV}]}{E[\mathrm{GV/m}]}\right)}\,, 
\end{split}
\end{equation}
where $\xi_i[\mathrm{eV}]$ is the potential energy of the bound electron,
$n^* \approx 3.69~\mathrm{Z}/\xi_i^{1/2}[\mathrm{eV}]$ is the effective principal 
quantum number, which depends on the ionization level $Z$, and $E[\mathrm{GV/m}]$ 
is the magnitude of electric field acting on the atom.
The electric field $E=E_{ion}$ for which the ionization rate becomes 
$W_{ADK}=0.1~\mathrm{fs^{-1}}$ 
is in this work considered as the ionization threshold.
In case of He $E_{ion}^{He}=93~\mathrm{GV/m}$ is predicted for the tunneling 
of the outer electron ($Z = 1,~\xi_i=23.6~\mathrm{eV}$), 
and for the inner one ($Z = 2,~\xi_i=54.4~\mathrm{eV}$) 
$E_{ion}^{He^+}=235~\mathrm{GV/m}$ applies.
The dynamics of electrons released by the above mechanism at a certain position in the wake 
can be addressed considering the electromagnetic Hamiltonian of a single electron 
$\mathcal{H}(\vec{x},\vec{P},t)=\sqrt{(mc^2)^2+c^2(\vec{P}+e\vec{A})^2}-e\Phi$,
characterized by the potentials $\Phi(\vec{x},t)$ and $\vec{A}(\vec{x},t)$ 
and the generalized coordinates $\vec{x}$ and $\vec{P}=\vec{p}-e\vec{A}$, 
where $\vec{p}$ is the momentum of the electron. 
In the co-moving system of reference  with $\zeta=z-v_{ph}t$ and
$v_{ph}$ the phase velocity of the wake, 
the electromagnetic potentials barely change over time compared
to their variation with $\zeta$. 
Therefore, the quasi-static approximation~\cite{sprangle1990} holds and 
$\partial_t=-v_{ph}\partial_\zeta=-v_{ph}\partial_z$.
In this case, provided that 
$\dot{\mathcal{H}} = \partial_t \mathcal{H} = -v_{ph}\partial_z \mathcal{H} = v_{ph} \dot{P_z}$,
the quantity $\mathcal{K} \equiv \mathcal{H} - v_{ph} P_z = mc^2 \gamma - v_{ph} p_z - e\Psi$ 
is a constant of motion~\cite{mora:217}. 
Here, we have defined the potential $\Psi \equiv \Phi-v_{ph} A_z$ 
related to the electric and magnetic fields by $E_z = -\partial_z \Psi$
and $E_r-v_{ph}B_{\phi} = -\partial_r \Psi$, while $\gamma$ is the Lorentz factor of the electron. 
Immediately after ionization, the electron has negligible energy and
can be considered at rest, thus $\mathcal{K}_i = mc^2 - e\Psi_i$. 
This electron is trapped into the wake if it follows a phase-space trajectory
such that its velocity $v$ reaches the velocity of the wake $v_{ph}$. 
When this happens, $\mathcal{K}_f = mc^2/\gamma_{ph} - e\Psi_f$, 
and since $\mathcal{K}_f=\mathcal{K}_i$, a trapping condition in terms
of the difference in potential between the initial $\Psi_i$ and trapped $\Psi_f$ 
positions is derived~\cite{Oz:2007mn}:
\begin{equation}
\label{eq:trap}
 \Delta\Psi \equiv \Psi_f - \Psi_i = -\frac{mc^2}{e}\left(1-\frac{1}{\gamma_{ph}}\right)
\end{equation}
For ultra-relativistic drivers ($v_{ph} \rightarrow c$ and $\gamma_{ph} \rightarrow \infty$),
Eq.~(\ref{eq:trap}) can be written as $\Delta\Psi = - mc^2/e$.
The initial position of the ionized electron inside the wake determines $\Psi_i$,
and consequently its final trapping position (if any) along the corresponding $\Psi_f$ 
equipotential contour.
The necessary trapping condition for electrons ionized inside the first wake period, 
ahead of the minimum of potential $\Psi_{min}$ at the rear of the ion cavity, 
is given by $\Psi_i > \Psi_{min} + mc^2/e \equiv \Psi_t$.  
The volume of injection is thus determined by the intersection of the volume of ionization
($E_i > E_{ion}$) with the volume satisfying the trapping criterion ($\Psi_{i} > \Psi_{t}$). 
Generally, the field configuration in the blow-out regime of PWFA  
can enable simultaneous ionization and trapping in two regions within the first wave 
bucket~\cite{Kirby:2009zzb}.
One is located at the driver beam position, where the radial electric field induces ionization.
This ionization region is sensitive to the oscillating behavior of the beam in the 
focusing ion-plasma column~\cite{clayton:2002} and fluctuations in the micro structure of 
its density profile.
The second region of ionization is located at the rear of the cavity, 
where the wakefields induce ionization. 
In contrast to the front, the fields at the back are stable in time
and barely dependent on details of the driver beam density profile and thus, 
provide a well-defined and controlled region for injection. 
The injection technique proposed in this work is designed
to inject electrons only from a narrow phase interval at the back of the cavity,
while preventing any contribution from the radial electric field of the driver.

\begin{figure}[!h]
 \centering
  \includegraphics[width=1.0\columnwidth]{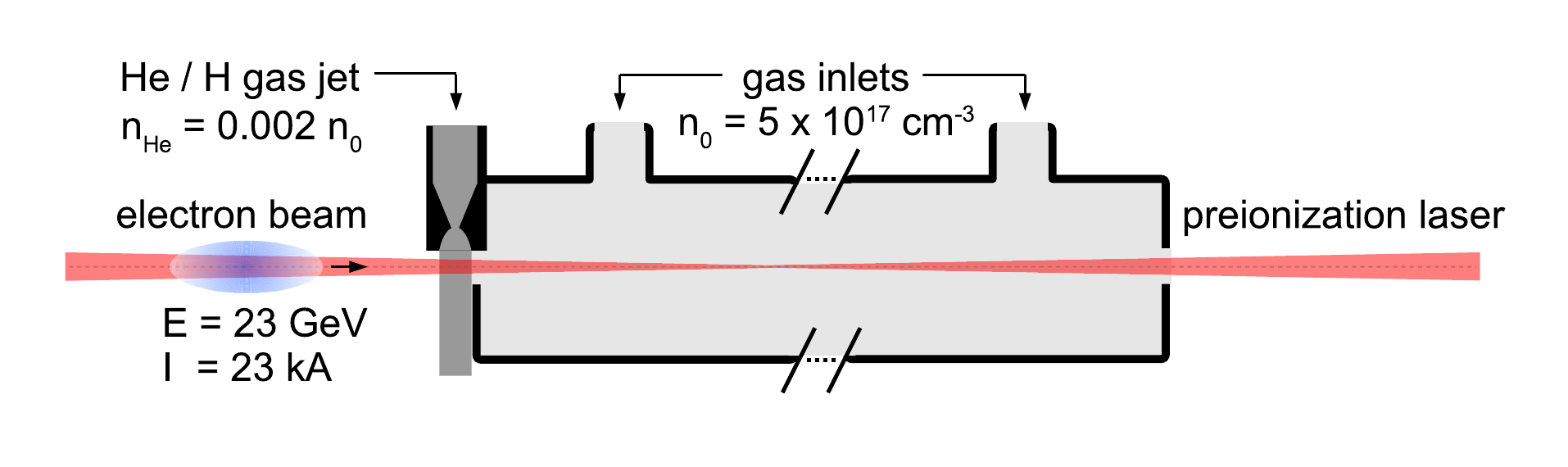}
  \caption{
    Schematic of the plasma-cell setup used in OSIRIS 3D simulations. 
    A thin jet column of neutral H/He gas mixture is immersed in a laser pre-ionized hydrogen plasma at $n_0$.
  }
\label{fig:plasmacell} 
\end{figure}

To illustrate this method, we consider in the following electron bunches 
similar to those provided by the FACET facility at SLAC.
These beams are approximated by Gaussian longitudinal ($\sigma_z = 14~\mathrm{\mu m}$) 
and transverse ($\sigma_x = \sigma_y = 10~\mathrm{\mu m}$) profiles 
with peak currents of $23~\mathrm{kA}$, transverse normalized emittances of 
$\epsilon_{x} = 50~\mathrm{\mu m}$ and $\epsilon_{y} = 5~\mathrm{\mu m}$,
and an energy of $23~\mathrm{GeV}$ with a relative spread of $1~\%$~\cite{Hogan:FACET}. 
The characteristics of these beams make them suitable to operate in the blow-out regime 
using a plasma with density $n_{0}=5\times10^{17}~\mathrm{cm^{-3}}$, 
which can be generated using current gas cell technology~\cite{Clayton:2010}.
As sketched in Fig.~\ref{fig:plasmacell}, a micro-nozzle~\cite{Hao:2005} 
fed by a hydrogen-helium mixture with tunable ratio and pressure, 
is positioned in the vicinity of the gas cell entrance. 
The gas jet emerging from the nozzle is spatially confined to a diameter of about 
$L_{He}=100~\mathrm{\mu m}$, 
forming a highly localized region in which helium is present while not mixing 
with the gas cell volume,
and overall maintaining a flat density distribution in line of sight of the electron beam. 
In order to prevent an excessive beam loading, 
we choose a He concentration of $n_{He} = 0.002~n_{0}$.
In this setup, the plasma can be pre-created by means of a laser coaxial to the electron beam
with an intensity $I_{L}\ge 1.52\times10^{14}~\mathrm{W/cm^2}$ capable to 
fully ionize the hydrogen, 
but not the helium at $I_{L}\ll1.14\times10^{15}~\mathrm{W/cm^2}$, 
where the limits for the laser intensities are calculated from 
$E_{ion}^{He}=93~\mathrm{GV/m}$ and $E_{ion}^{H}=34~\mathrm{GV/m}$, 
the ionization thresholds for He and H respectively. 

\begin{figure}[!t]
 \centering
  \includegraphics[width=\columnwidth]{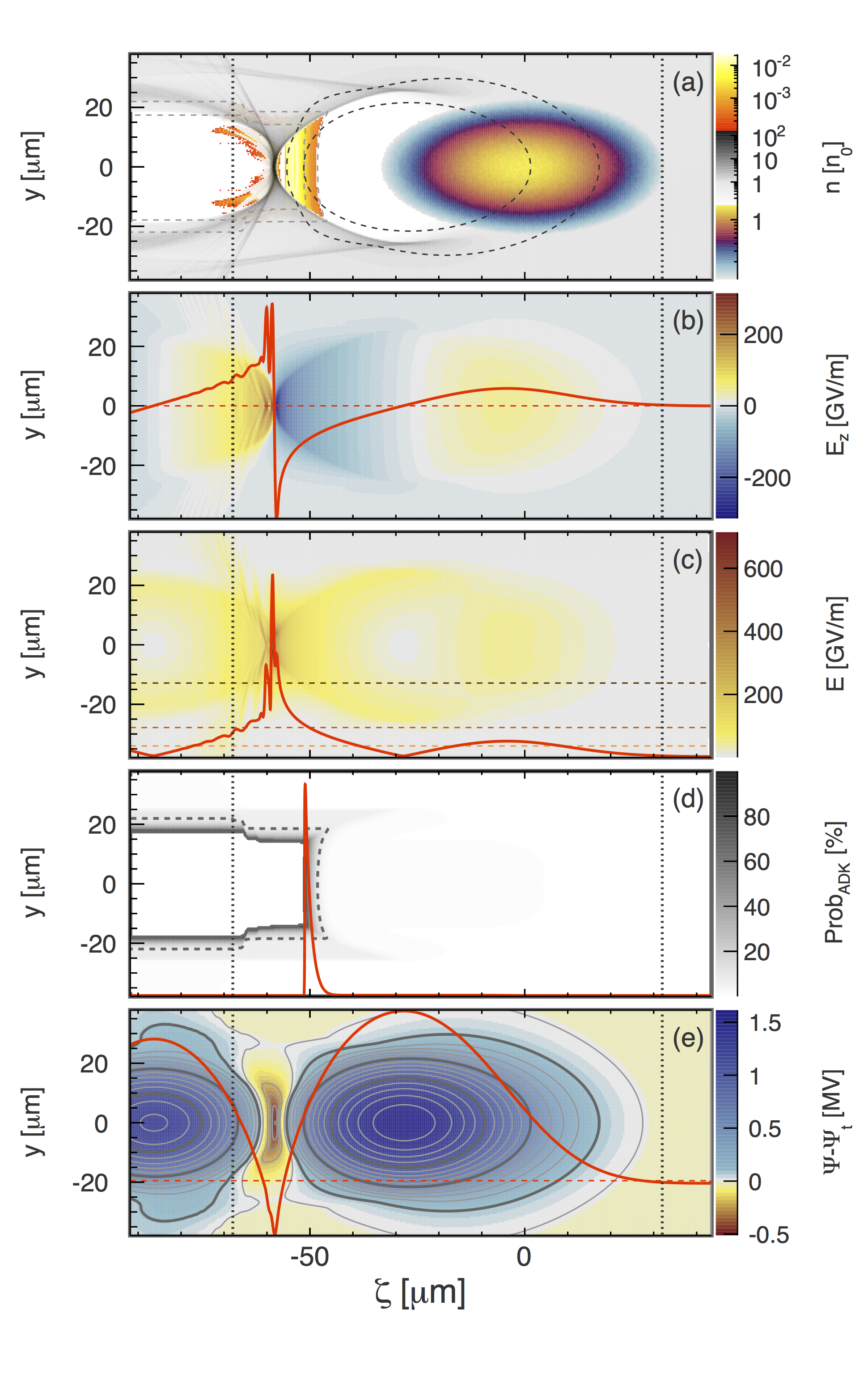}
  \caption {
   Central slices from a 3D OSIRIS simulation depicting the trapping conditions and showing ionization of electrons from a He dopant by wakefields 
    excited by a FACET-type beam in a pre-created uniform plasma with a density of $n_e=5\times10^{17}~\mathrm{cm^{-3}}$. 
    (a) Spatial particle density. Plasma (gray palette), driver beam (blue-yellow palette), and electrons ionized from He (red-yellow palette).
    (b) Longitudinal wakefields (blue-red palette) and on-axis (red line).
    (c) Total electric field (red palette) and on-axis values (red line). 
Light (dark) dotted red line shows the ionization threshold of the outer (inner) He electron.
    (d) ADK ionization probability of the first level of He (gray palette), 
        the contour at $10~\%$ probability (dotted line), and the on-axis values (red line).
    (e) The electric potential $\Psi-\Psi_t$ (blue-red palette), its contours in steps of $\Delta\Psi = 0.2\times(mc^2/e)$, 
        and the on-axis values (red line).
Vertical dotted lines show the limits of the He column.
  }
\label{fig:rakesnap} 
\end{figure}

Three-dimensional (3D) simulations of this setup have been performed
using the particle-in-cell (PIC) code OSIRIS~\cite{Fonseca:osiris}, 
which is capable of emulating ionization effects using the ADK model~\cite{Ammosov:1986adk}.
The moving window simulation box dimensions are $18\times30\times30~k_p^{-3}$ with a cell size of $0.036\times0.060\times0.060~k_p^{-3}$.
Fig.~\ref{fig:rakesnap}(a) shows the electron density of the plasma (gray color palette), 
the driver beam (blue-yellow palette) and the ionized electrons (red-yellow palette)
in the central slice ($y~\mathrm{vs.}~\zeta\mathrm{, at }~x=0$) of the simulation 
inside the He region. 
The FACET beam drives accelerating wakefields which exhibit peak values greater than 
$200~\mathrm{GV/m}$ (Fig.~\ref{fig:rakesnap}(b)).
The magnitude of the electric field~(Fig.~\ref{fig:rakesnap}(c)) 
in the accelerating region of the wake exceeds by far 
the threshold for the ionization of helium $E_{ion}^{He}$, 
whereas in the decelerating region of the wake it is significantly lower.
To restrict the area of high ionization rate to the accelerating phase, 
the radial space charge field of the driver $|E_r|$, which is inversely dependent on the transverse beam size, 
must be less than $E_{ion}^{He}$ during passage through the He gas jet.
This can be ensured by placing the jet at the entrance of the plasma target,
well before the beam experiments its first compression induced by the focusing ion-plasma cavity~\cite{clayton:2002}. 
The length scale of transverse focusing of an unmatched beam is given by the betatron wavelength 
$\lambda_{\beta}=\sqrt{2\gamma}~\lambda_p$~\cite{Esarey:2002beta}, 
in the considered case $\lambda_\beta \approx 14~\mathrm{mm}$. 
Fig.~\ref{fig:rakesnap}(d) shows the probability of ionization $P_{ADK}(r,\zeta)$ 
of the He atoms streaming backwards with respect to the wake, 
obtained by integrating the ionization rate $W_{ADK}$ in Eq.~(\ref{eq:ionization}) along $\zeta$. 
The contours where $P_{ADK}$ reaches $10\%$ and $100\%$ of ionization are drawn in gray dotted 
and bold lines, respectively, defining a narrow phase interval $\{\zeta_{10},\zeta_{100}\}$
extending up to the borders of the bubble, from which $90\%$ 
of all possibly trapped He electrons will be emerging 
($\Delta\zeta_{ion} \equiv \zeta_{10} - \zeta_{100} \approx 3~\mathrm{\mu m}$).
Fig.~\ref{fig:rakesnap}(e) depicts the trapping potential $\Psi-\Psi_t$, 
where positive values correspond to regions which allow trapping ($\Psi>\Psi_t$) 
and where equipotential contours are shown in steps of $0.2\times mc^2/e$. 
The intersection of the volume with high ionization probability 
and the volume allowing trapping yields the volume from which injected electrons can originate~(Fig.~\ref{fig:rakesnap}(a)). 
However, trapping is also affected by the transverse dynamics of the electron in the plasma wave,
i.e.~by its initial radial position. 
Electrons released close to the boundary of the cavity may escape 
before the focusing force pushes them towards a stopping contour near the axis.
A sufficient condition for trapping of an electron with a given initial radius
is that it was trapped even if it kept the same radius when falling back 
with respect to the plasma wave, 
i.e. it reaches $\Psi_f$ before the bubble boundary in straight backwards propagation.
In this example, the maximum radius fulfilling the above condition is
$R_{max}\approx 12~\mathrm{\mu m}$ (c.f.~Fig.~\ref{fig:rakesnap}(a)).
This allows for an estimation of the actual volume of injection 
$V_{inj} \simeq \pi R_{max}^2~\Delta\zeta_{ion}$ and hence for the total trapped charge 
$Q_{He} \simeq -e n_{He}~\pi R_{max}^2~L_{He} = 7.2~\mathrm{pC}$ during passage through the He-doped gas column. 

\begin{figure}[!t]
 \centering
  \includegraphics[width=\columnwidth]{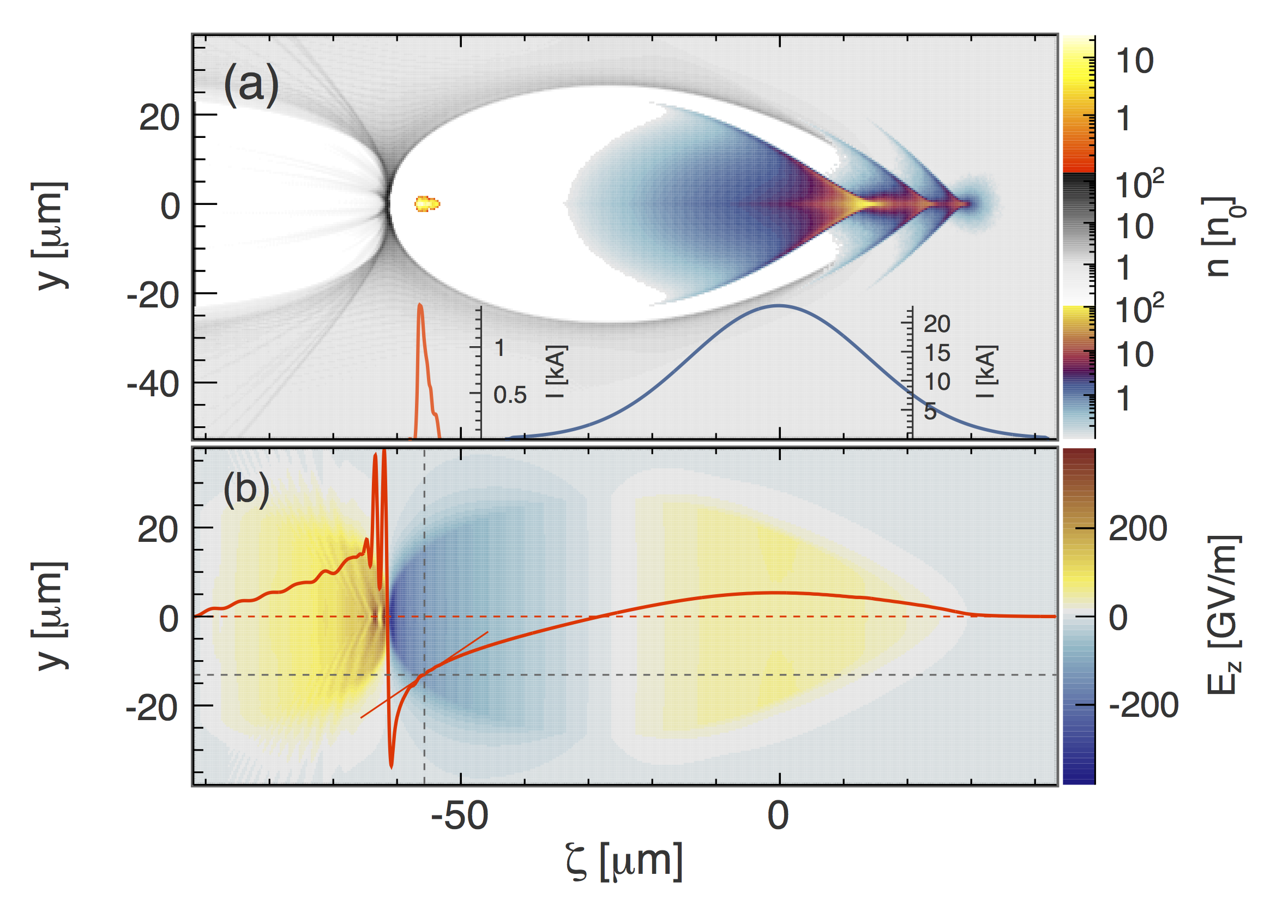}
  \caption {
    PIC simulation after $20~\mathrm{mm}$ of beam propagation.
    (a) Charge densities of the electron plasma, beam and injected bunch. 
    The curves show currents of the the drive beam (blue) and injected electrons (orange). 
    (b) Longitudinal wakefields.
  }
\label{fig:rakesnap2} 
\end{figure}

Fig.~\ref{fig:rakesnap2}(a) shows a short ($0.8~\mathrm{\mu m}$ rms) bunch of electrons 
injected from the neutral He by means of the wakefields, in the above discussed simulation.
With a total charge of $8.8~\mathrm{pC}$ and a maximum peak current of $1.5~\mathrm{kA}$,
the injected beam has been accelerating for $20~\mathrm{mm}$,
positioned at $\langle\zeta_f\rangle=-55.7~\mathrm{\mu m}$,
where the longitudinal electric field is $E_z(\langle\zeta_f\rangle) \approx 130~\mathrm{GV/m}$
(Fig.~\ref{fig:rakesnap2}(b)). 
Most physical properties of the trapped bunch can be estimated 
from the initial phase-space distribution. 
Trapped electrons with the same initial value of $\Psi_{i}$ are positioned  
approximately on the same co-moving phase near axis during acceleration, 
fulfilling $\Psi_f(\zeta_f) = \Psi_i(\zeta_{i}) - mc^2/e$, 
and thus will be accelerated by the same field value $E_z(\zeta_f)$. 
However, each one of these slices in $\zeta_{f}$ is composed of electrons ionized at different longitudinal
positions along the He column, and therefore accelerated at different times, 
producing a finite spread in longitudinal momentum in every slice given by 
$\Delta p_{z}(\zeta_f) \simeq - e E_{z}(\zeta_f)~L_{He}$,
which, at the average position of the bunch gives 
$\Delta p_{z}(\langle\zeta_f\rangle) \approx 13~\mathrm{MeV}$.  
Moreover, the total relative energy spread is proportional to the variation of $E_z$ along
the bunch length, which in case of a negligible beam loading and sufficiently short bunches, 
is approximately given by 
$\Delta\gamma/\gamma \simeq \partial_{\zeta}E_z(\langle\zeta_f\rangle)/E_z(\langle\zeta_f\rangle)~\Delta\zeta_f$. 
From Fig.~\ref{fig:rakesnap2}(b), $\partial_{\zeta}E_z(\langle\zeta_f\rangle) \approx 10~\mathrm{(GV/m)\,\mu m^{-1}}$, 
and $\Delta\gamma/\gamma \approx 6\mathrm{\%}$.
Electrons belonging to the same $\zeta_f$ slice come from different radial positions 
along their initial $\Psi_i$ contour.
Assuming full decoherence for every slice, an upper estimate
of the uncorrelated normalized transverse emittance 
$\epsilon_{y}=\sqrt{\langle y^2 \rangle\langle p_{y}^2 \rangle-\langle yp_y\rangle^2}/mc$,
can be given in terms of the initial transverse extend of the slice~\cite{Kirby:2009zzb}
$\epsilon_{y} = k_p \langle y_i^2 \rangle/4$.
Considering for simplicity, the largest $\Psi_i$ contour to be uniformly distributed up to $R_{max}$,
the estimated maximum sliced emittance yields 
$\epsilon_{y,max} = k_p R_{max}^2/12 \approx 1.4~\mathrm{\mu m}$. 

\begin{figure}[!h]
 \centering
  \includegraphics[width=1.0\columnwidth]{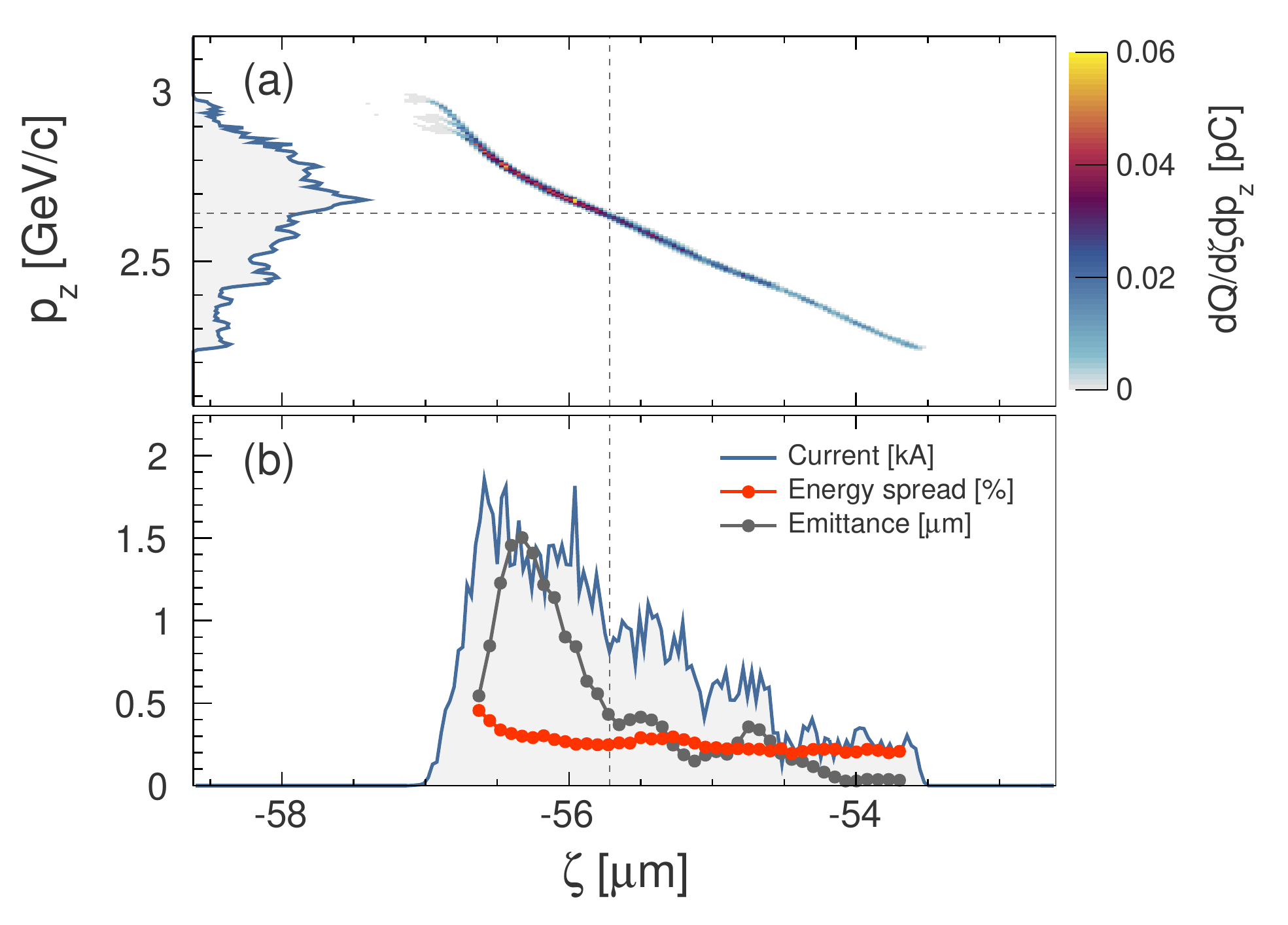}
  \caption{
    Witness bunch properties after $20~\mathrm{mm}$ of acceleration.
    (a) shows the electron distribution of the bunch in longitudinal phase space ($p_{z}$ vs. $\zeta$ plane).
    The projection of this distribution in $p_{z}$ is depicted on the left axis.
    (b) displays the bunch current in dependence of the co-moving coordinate $\zeta$.
    The relative energy spread (red points) and the transverse emittance (gray points) 
    are plotted for different longitudinal slices along the bunch.   
  }
\label{fig:bunchprop} 
\end{figure}

The properties of the simulated injected bunch after $20~\mathrm{mm}$ of acceleration 
are summarized in Fig.~\ref{fig:bunchprop}.
The longitudinal phase-space (Fig.~\ref{fig:bunchprop}(a)) exhibits linear chirp 
with an average energy of $\sim2.6~\mathrm{GeV}$ and a total relative energy spread of~$6\%$.
The sliced bunch properties can be seen in more detail in Fig.~\ref{fig:bunchprop}(b).
The current profile has a maximum at the tail of the bunch of $\sim1.5~\mathrm{kA}$ 
and linearly decays towards its front~(Fig.~\ref{fig:bunchprop}(b)).
The relative energy spread ($\sim0.3\%$), and the normalized transverse emittance ($\le 1.5~\mathrm{\mu m}$)
are shown for different slices in $\zeta$, demonstrating an excellent agreement with the 
analytical estimations given previously. 

In summary, a new strategy for the injection of electrons in PWFA is proposed and demonstrated 
using 3D PIC simulations. 
The described method leads to a controlled ionization-induced self-injection of electrons into 
blow-out plasma wakes in a simple experimental setup, 
which utilizes only the wakefields at the rear of the ion cavity to trigger the injection and trapping
of electrons from a neutral atomic species into a well-defined phase of the plasma wake.  
As a result, high-quality electron bunches can be produced with short pulse lengths ($\le 1~\mathrm{\mu m}$),
low normalized emittances ($\sim 1~\mathrm{\mu m}$), and low uncorrelated energy spread ($< 1~\%$)
on a GeV-energy scale.
The first experiments demonstrating such beam quality will be regarded as important milestones 
in the ongoing endeavor to advance plasma-based particle accelerators for their future application 
in photon science and high-energy physics.

We thank the OSIRIS consortium (IST/UCLA) for access to the OSIRIS code. 
Special thanks for support go to J. Vieira and R. Fonseca. 
Furthermore, we acknowledge the grant of computing time by the J\"{u}lich Supercomputing Centre on JUQUEEN under Project No. HHH09. 
We would like to thank DESY IT for their support concerning simulations and data storage at DESY 
and the Humboldt Foundation for financial support.

\end{document}